# Tuneable Sieving of Ions Using Graphene Oxide Membranes

J. Abraham[1,2,3], K. S. Vasu[1,2], C. D. Williams[2], K. Gopinadhan[3], Y. Su[1,2], C. Cherian[1,2], J. Dix[2], E. Prestat[4], S. J. Haigh[4], I. V. Grigorieva[1], P. Carbone[2], A. K. Geim[3] & R. R. Nair[1,2*]

[1]National Graphene Institute, University of Manchester, Manchester, M13 9PL, UK.

[2]School of Chemical Engineering and Analytical Science, University of Manchester, Manchester, M13 9PL, UK.

[3]School of Physics and Astronomy, University of Manchester, Manchester M13 9PL, UK.

[4]School of Materials, University of Manchester, Manchester, M13 9PL, UK.

rahul@manchester.ac.uk

**Graphene oxide membranes show exceptional molecular permeation properties, with a promise for many applications. However, their use in ion sieving and desalination technologies is limited by a permeation cutoff of ~9 Å, which is larger than hydrated ion diameters for common salts. The cutoff is determined by the interlayer spacing $d$ ~13.5 Å, typical for graphene oxide laminates that swell in water. Achieving smaller $d$ for the laminates immersed in water has proved to be a challenge. Here we describe how to control $d$ by physical confinement and achieve accurate and tuneable ion sieving. Membranes with $d$ from ~ 9.8 Å to 6.4 Å are demonstrated, providing the sieve size smaller than typical ions' hydrated diameters. In this regime, ion permeation is found to be thermally activated with energy barriers of ~10–100 kJ/mol depending on $d$. Importantly, permeation rates decrease exponentially with decreasing the sieve size but water transport is weakly affected (by a factor of <2). The latter is attributed to a low barrier for water molecules entry and large slip lengths inside graphene capillaries. Building on these findings, we demonstrate a simple scalable method to obtain graphene-based membranes with limited swelling, which exhibit 97% rejection for NaCl.**

Selectively permeable membranes with sub-nm pores attract strong interest due to analogies with biological membranes and potential applications in water filtration, molecular separation and desalination[1-8]. Nanopores with sizes comparable to, or smaller than, the diameter $D$ of hydrated ions are predicted to show enhanced ion selectivity[7,9-12] because of dehydration required to pass through such atomic-scale sieves. Despite extensive research on ion dehydration effects[3,7,9-13], experimental investigation of the ion sieving controlled by dehydration has been limited because of difficulties in fabricating uniform membranes with well-defined sub-nm pores. The realisation of membranes with dehydration-assisted selectivity would be a significant step forward. So far, research into novel membranes has mostly focused on improving the water flux rather than ion selectivity. On the other hand, modelling of practically relevant filtration processes shows that an increase in water permeation rates above the rates currently achieved (2-3 L/m$^2$×h×bar) would not contribute greatly to the overall efficiency of desalination[8,14,15]. Alternative approaches based on higher water-ion selectivity may open new possibilities for improving filtration technologies, as the performance of state-of-the-art membranes is currently limited by the solution-diffusion mechanism, in which water molecules dissolve in the membrane material and then diffuses across the membrane[8]. Recently, carbon nanomaterials including carbon nanotubes (CNT) and graphene have emerged as promising membrane materials. Unfortunately, such membranes are also difficult to manufacture on industrial scale[4,5,8,16]. In particular, monolayer graphene was suggested as a membrane for ion exclusion by creating sub-nm pores using ion bombardment and selective etching[1-4]. However, it is difficult to achieve the high density and uniformity of such pores, which is required for industrial applications, because of the stochastic nature of the involved processes. In contrast, graphene oxide (GO), a chemical derivative of graphene with oxygen functionalities[17], has attracted wide-spread interest due to its exceptional water permeation and molecular sieving properties[18-20] as well as realistic prospects for industrial scale production[21,22]. Molecular permeation through GO membranes is believed to occur along a network of pristine graphene channels that develop between functionalized areas of GO sheets[18] (typically, an area of 40-60% remains free from functionalization[23,24]), and their sieving properties are defined by the interlayer spacing, $d$, which depends on the humidity of the surroundings[18,19]. Immersing GO membranes in liquid water leads to intercalation of 2–3 layers of water molecules between individual GO sheets, which results in swelling and $d \approx 13.5$ Å. The effective pore-size of 9 Å in these swollen membranes (excluding the space occupied by carbon atoms) is larger than a typical size of hydrated ions and restricts possible uses of GO for size-exclusion based ion sieving[19]. A number of strategies have been tried to inhibit the swelling effect, including partial reduction of GO[25], ultraviolet reduction of GO-titania hybrid membranes[26], and covalent crosslinking[27-29]. In this report, we investigate ion permeation through GO laminates with $d$ controlled from $\approx 9.8$ to $\approx 6.4$ Å, which is achieved by physical confinement (Fig. 1a). Our results show that the changes in $d$ dramatically alter ion selectivity due to dehydration effects whereas permeation of water molecules remains largely unaffected.

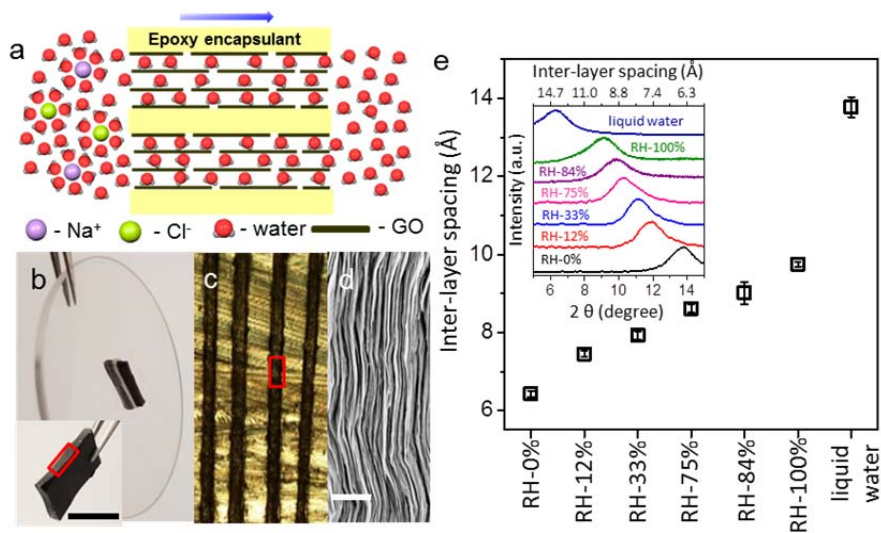

**Figure 1| Physically confined GO membranes with tuneable interlayer spacing.** (**a**) Schematic illustrating the direction of ion/water permeation along graphene planes. (**b**) Photograph of a PCGO membrane glued into a rectangular slot within a plastic disk of 5 cm in diameter. Inset: Photo of the PCGO stack before it was placed inside the slot. Scale bar, 5 mm. (**c**) Optical micrograph of the cross-sectional area marked by a red rectangle in (b), which shows 100-μm-thick GO laminates (black) embedded in epoxy. The latter is seen in light yellow with dark streaks because of surface scratches. (**d**) SEM image from the marked region in (c). Scale bar, 1μm. (**e**) Humidity dependent $d$ found using X-ray diffraction (inset). The case of liquid water is also shown. Error bars: Standard deviations using at least two measurements from three different samples.

Thick (≈100 μm) GO laminates were prepared by vacuum filtration of aqueous GO solutions, as reported previously[18]. The laminates were cut into rectangular strips (4 mm × 10 mm) and stored for one to two weeks at different relative humidities (RH), achieved using saturated salt solutions[18,30]. The resulting interlayer spacing was measured by X-ray diffraction as shown in Fig. 1e and varied from ≈ 6.4 to 9.8 Å with RH changing from zero to 100%. GO laminates soaked in liquid water showed $d ≈ 13.7 ± 0.3$ Å. All these values agree with previous reports, where the changes in $d$ were attributed to successive incorporation of water molecules into various sites between GO sheets[31]. Individual GO strips with desirable $d$ were then encapsulated and stacked together using Stycast epoxy as shown in Figs. 1b,c to increase the available cross-section for filtration to ~1 mm (see Methods and supplementary Fig. S1). The stacked GO laminates, now embedded in the epoxy (Fig. 1c), are referred to as physically confined GO (PCGO) membranes because the epoxy mechanically restricts the laminate's swelling upon exposure to RH or liquid water (Methods). The stacks were glued into a slot made in either metal or plastic plate (Fig. 1b). Two sides of these stacked PCGO membranes were then trimmed off to make sure that all the nanochannels are open (Fig. 1d) before carrying out permeation experiments, in which ions and water molecules permeates along the lamination direction as shown in Fig. 1a.

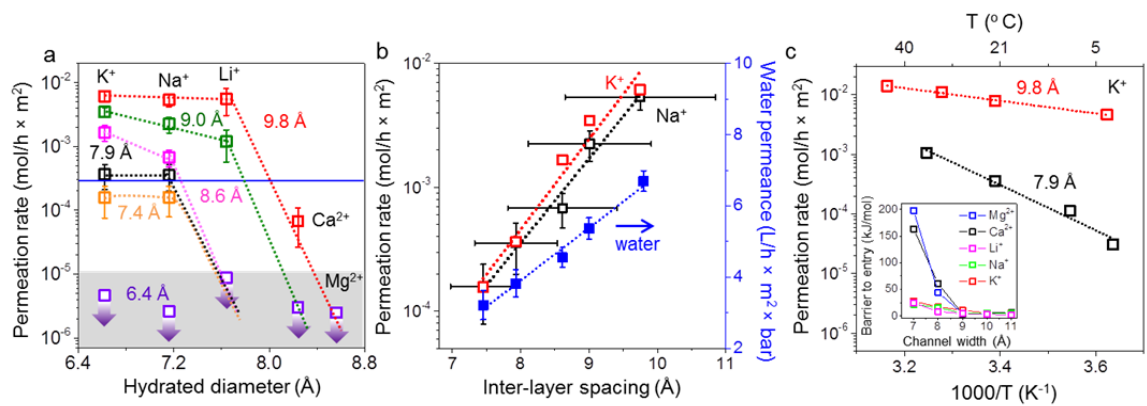

**Figure 2| Tuneable ion sieving.** (**a**) Permeation rates through PCGO membranes with different interlayer distances (colour coded). The salts used: KCl, NaCl, LiCl, CaCl$_2$ and MgCl$_2$. The hydrated diameters are taken from Ref. 32 (supplementary section 4). Dashed lines: Guides to the eye indicating a rapid cutoff in salt permeation, which is dependent on $d$. Grey area: Below-detection limit for our measurements lasting 5 days, with arrows indicating the limits for individual salts. The horizontal blue line indicates our detection limit for Cl$^-$. Above the latter limit, we found that both cations and anions permeated in stoichiometric quantities. Error bars: Standard deviation. (**b**) Permeation rates for K$^+$ and Na$^+$ depend exponentially on the interlayer distance (left axis). Water permeation varied only linearly with $d$ (blue squares, right axis). The dotted lines are best fits. The horizontal error bars correspond to a half-width for the diffractions peaks in Fig. 1e and are same for all the three data sets. The vertical error bars indicate the standard deviation. The errors for K$^+$ are similar to those for Na$^+$ and omitted for clarity. (**c**) Temperature dependence for K$^+$ permeation. Dotted lines: Best fits to the Arrhenius behaviour. Inset: Energy barriers for various ions and different $d$, as found in our molecular dynamics simulations.

Our measurement setup was similar to the one previously reported[19] and consisted of two Teflon compartments (feed and permeate) separated by a PCGO membrane (supplementary Fig. S2). The feed and permeate compartments were filled with 10 mL of a salt solution and deionized water, respectively. Quantitative analysis of anion and cation permeation between the compartments was carried out using ion chromatography (IC) and inductively coupled plasma atomic emission spectroscopy (ICP-AES), respectively. As expected, the ion concentration in the permeate compartment increases with time and with increasing the concentration of the feed solution (supplementary section 3 and supplementary Fig. S3). Fig. 2a summarises our results obtained for various ions permeating through PCGO membranes with different interlayer spacing. One can see that the permeation rates and the cutoff diameter for salt permeation decrease monotonically with decreasing $d$. Membranes with $d \approx$ 6.4 Å showed no detectable ion concentration in the permeate even after five days. This further confirms that our PCGO membranes do not swell in water over time, despite a finite mechanical rigidity of the epoxy confinement. When plotted as a function of $d$, the observed ion permeation rates for $Na^+$ and $K^+$ showed an exponential dependence, decreasing by two orders of magnitude as $d$ decreased from 9.8 to 7.4 Å (Fig. 2b). In contrast, the same PCGO membranes (supplementary section 5) showed only a little variation in permeation rates for water (Fig. 2b), decreasing by a factor of $\approx$ 2 for the same range of $d$. We note that this observation also rules out that the exponential changes in ion permeation could be related to partial clogging of graphene capillaries.

Both the observed relatively high permeation rates for $Li^+$, $K^+$ and $Na^+$ for d >9 Å and their exponential decay for smaller $d$ are surprising. Indeed, when considering steric (size-exclusion) effects, it is often assumed that ions in water occupy a rigid volume given by their hydrated diameters $D$. If this simplification was accurate, our PCGO membranes should not allow permeation of any common salt. Indeed, the effective pore size $\delta$ can be estimated as $d - a$, where $a \approx 3.4$ Å is the thickness of graphene[18,33]. This yields $\delta \approx 6.4$ Å for our largest capillaries ($d \approx 9.8$ Å), which is smaller than $D$ for all the ions in Fig. 2a. This clearly indicates that ion sieving is not purely a geometric effect. On the other hand, if we assume that hydrated ions do fit into the nanochannels and their permeation is only limited by diffusion through water, the expected permeation rates should be significantly higher than those observed experimentally. For classical diffusion the permeation rate $J$ is given by

$$J = Diff \times \Delta C \times A_{eff}/L \qquad (1)$$

where $\Delta C$ is the concentration gradient across the membrane (1 M for the experiments in Fig. 2), $A_{eff}$ the total cross-sectional area of nanocapillaries ($\approx$ 3-8 mm$^2$), $L$ the diffusion length through the PCGO membrane ($\approx$ 3 mm) and $Diff$ is the diffusion coefficient for ions in water (typically, $Diff \sim 10^{-5}$ cm$^2$/s; see supplementary section 6). Eq. (1) yields rates that are 2 to 4 orders magnitude higher than those shown in Fig. 2. This is in stark contrast to the sieving properties of GO laminates with $d \approx 13.5$ Å which showed an enhancement rather than suppression of ion diffusion[19]. Clearly, the fact that the available space $\delta$ in PCGO laminates becomes smaller than $D$ pushes the permeating hydrated ions into a new regime, distinct both from ions moving through wider nanocapillaries and from permeation behaviour of pure water. In the latter case, as shown in Fig. 2b, permeation rates for water molecules (whose size is smaller than $\delta$) are 3 orders of magnitude higher than those estimated from the standard Hagen-Poiseuille equation using non-slip boundary conditions and the given dimensions of nanocapillaries (supplementary section 5). Similar flow enhancement (by 3 order of magnitude) was recently reported for artificial graphene capillaries and attributed to a large slip length of ~60 nm for water on graphene[33].

To gain an insight into the mechanism of ion permeation through our membranes, we carried out permeation experiments at different temperatures, $T$ (Fig. 2c). For both channel sizes, $d$ = 9.8 and d= 7.9 Å, the permeation rates follow the Arrhenius equation, $\exp(-E/k_B T)$, i.e., show activation behaviour. Here $E$ is the energy barrier and $k_B$ the Boltzmann constant. The data yield $E = 72 \pm 7$ and $20 \pm 2$ kJ/mol for K$^+$ ion permeation through PCGO membranes with $d \approx 7.9$ and 9.8 Å, respectively. The exponential dependence explains the fact that the observed ion diffusion rates are orders of magnitude smaller than those given by eq. (1), as at room temperature $E>>k_BT$ for both channel sizes. The activation behaviour is also in agreement with recent theoretical predictions that nanopores with diameters < 10 Å should exhibit significant energy barriers because of the required partial dehydration for ion's entry[3,7,9-12]. Qualitatively, this mechanism can be explained as follows. In a bulk solution, water molecules stabilize ions by forming concentric hydration shells. For an ion to enter a channel with $\delta < D$, some water molecules must be removed from the hydration shell. The higher the ion charge, the stronger it attracts water molecules. Accordingly, ions with larger hydration free energies and, therefore, 'tougher' water shells are expected to experience larger barriers for entry into atomic-scale capillaries and exponentially smaller permeation rates. Ions with weakly bound shells are easier to strip from their water molecules and allow entry into nanochannels. Similar arguments can be used to understand why water does not exhibit any exponential dependence on $d$: Water-water interactions are weak, so that it costs relatively little energy to remove surrounding water from water molecules entering the capillaries[10].

To support the proposed mechanism of dehydration-limited ion permeation for our PCGO membranes, we employ the previously suggested model of a network of graphene capillaries, which was developed to account for the fast permeation of water through GO membranes[12,18,19]. Within this model, we performed molecular dynamics simulations to find energy barriers for various ions entering graphene capillaries of different widths (supplementary section 6). As seen in Fig. 2c the energy barrier $E$ exhibits a sharp increase for $d < 9$ Å and is considerably larger for divalent ions compared to monovalent ones, in agreement with our experiments and the above discussion (Fig. 2a). Quantitatively, the obtained $E$ are of the same order of magnitude as those found experimentally; the discrepancy in exact values can be expected because realistic GO channels contain non-stoichiometric functionalities, rough edges, etc. which are difficult to model accurately. We also performed simulations to evaluate a possible contribution of diffusion rates through capillaries themselves into the overall permeation rates. Our results show that the diffusion coefficient for K$^+$ changes with $d$ but the effect is small compared to the exponential decrease in permeation rates, which was observed experimentally (supplementary section 6). This suggests that the energy barrier associated with dehydration is the dominant effect in our case of sub-nm capillaries.

Finally, the exponential suppression of ion permeation combined with fast water transport in PCGO membranes make them an interesting candidate for water filtration applications. Even though scalable production of such membranes is difficult, one can envisage alternative fabrication techniques to control $d$ in GO laminates are required. To this end, we show that it is possible to restrict the swelling of GO membranes in liquid water, for example, simply by incorporating graphene flakes into GO laminates (see supplementary section 7). The resulting composites referred to as GO-Gr membranes exhibit notably less swelling (difference in $d$ of $\approx 4$ Å) with respect to the standard GO laminates (Fig. 3a). The observed large difference in $d$ can be due to graphene's hydrophobicity that limits the water intake. The ion permeation rate through GO-Gr membranes was found to be suppressed by more than two orders of magnitude compared to GO (Fig. 3b), in agreement with the projected rates for the given extent of swelling if the exponential dependence of Fig. 2 is extrapolated. At the same time, water permeation rates are essentially unaffected by the incorporation of graphene into GO laminates (decrease only by 20%; supplementary section 7). The salt rejection properties of our GO-Gr membranes were further investigated using forward osmosis, where we employed concentrated (3 M) sugar and (0.1M) NaCl solutions as the draw and feed solutions, respectively (see supplementary section 7). Salt rejection was calculated as 1-$C_d$/$C_f$ where $C_d$ and $C_f$ are the concentration of NaCl at the draw and feed sides, respectively. Our analysis yielded ≈ 97% salt rejection for the GO-Gr membranes with a water flux of ≈ 0.5 L/m$^2$×h. Even though the flux is lower than 5-10 L/m$^2$×h typical for forward osmosis[35], we believe this characteristic can be significantly improved by decreasing the membrane thickness to 1 μm or less (see supplementary section 7). Such thicknesses are readily achievable for GO laminates[20] and can result in fluxes >5 L/m$^2$×h.

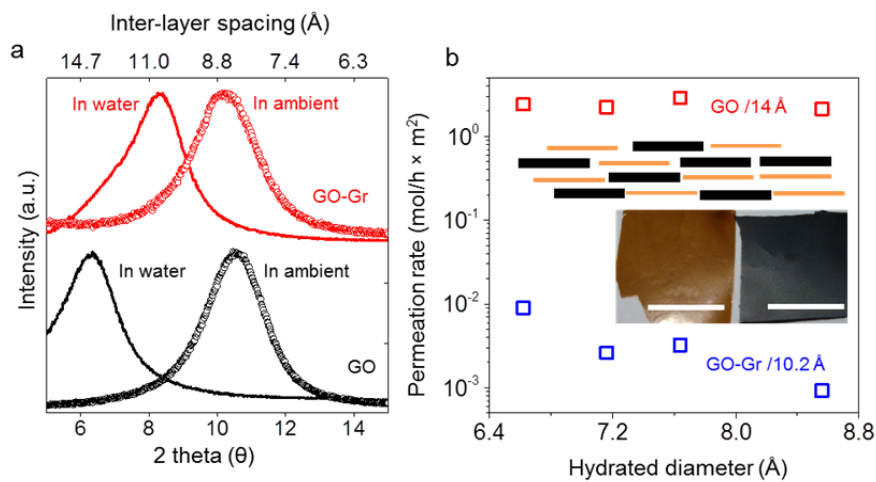

**Figure 3| GO membrane with limited swelling.** (**a**) X-ray diffraction showing shifts of the (001) peak due to swelling in liquid water for the standard GO laminate and a composite made from graphene and graphene oxide. (**b**) Ion permeation rates (same salts as in Fig.2) through such GO and GO-Gr membranes with a thickness of 5 μm. Top inset: Schematic of the GO-Gr structure (brown blocks – GO, black – graphene). Bottom inset: Photographs of GO and GO-Gr membranes (left, and right, respectively). Scale bars, 1 cm.

In conclusion, we have demonstrated the possibility to control the interlayer spacing in GO membranes in the range below 10 Å. In this regime the capillary size is smaller than hydrated diameters of ions and their permeation is exponentially suppressed with decreasing $d$. The suppression mechanism can be described in terms of additional energy barriers that arise because of the necessity to partially strip ions from their hydrated shells so that they can fit inside the capillaries. Water transport is much less affected by $d$. Our work shows a possible route to production of GO membranes with controllable interlayer spacing for desalination applications.

**Methods**

**Preparation of GO membranes.** The aqueous suspension of graphene oxide (GO) was prepared by dispersing millimeter sized graphite oxide flakes (purchased from BGT Materials Limited) in distilled water using bath sonication for 15 hours. The resulting dispersion was centrifuged 6 times at 8000 rpm to remove the multilayer GO flakes. Subsequently, free standing GO membranes of thickness ≈ 100 μm were prepared by vacuum filtration of supernatant GO suspension[19] through an Anodisc alumina membrane filter (0.2 μm pore size and a diameter of 47 mm, purchased from Millipore). As-prepared GO membranes were dried in an oven for 10 hours at 45 °C and cut into rectangular strips of dimension of 4 mm×10 mm (Supplementary Fig. S1).

**Tuning interlayer spacing in GO laminates.** GO membranes with different IL spacing were prepared by storing them in a sealed container with different RH of 0%, 12%, 33%, 75%, 84% and 100%. To this end, we used saturated solutions of LiCl (12% RH), $MgCl_2$ (33%), NaCl (75%) and KCl (84%), which were prepared by dissolving excess amounts of salts in deionised water[30,36]. A humidity meter was used inside the container to check that the salts provided the literature values of RH. As a zero humidity environment, we used a glove box filled with Ar and $H_2O$ content below 0.5 ppm. 100% RH was achieved inside a sealed plastic container filled with a saturated water vapour at room $T$.

**Analysis of the interlayer spacing.** X-ray diffraction (XRD) measurements in the $2\theta$ range of 5° to 15° (with a step size of 0.02° and recording rate of 0.1 s) were performed using a Bruker D8 diffractometer with Cu K$\alpha$ radiation ($\lambda = 1.5406$ Å). To collect an XRD spectrum from a GO membrane stored at a specific RH, we have created the same humid environment inside a specimen holder (Bruker, C79298A3244D83/85) and sealed it with the GO membrane. For the case of zero humidity, an airtight sample holder (Bruker, A100B36/B37) was used. All spectra were taken with a short scanning time to avoid possible hydration/dehydration of the GO membranes. From XRD analysis of the (001) reflection, IL $d$ for 0%, 12%, 33%, 75%, 84% and 100% RH are found to be 6.4, 7.4, 7.9, 8.6, 9 and 9.8 Å respectively.

**Fabrication of PCGO membranes.** After achieving the desired $d$ by using different humidity, each rectangular strip was immediately glued and stacked with Stycast 1266. This stack was then immediately transferred to the same humid environment (where the GO laminates were initially stored) for curing the epoxy overnight. Finally, the resulting stacks were glued into a slot in a plastic or copper plate as shown in Fig. 1. An epoxy layer present at the top and bottom cross sections of the glued stacks was carefully cleaved to produce a clean surface for permeation experiments. The cleaved cross-section was also checked under an optical microscope to remove any possible epoxy residues. The entire fabrication procedure is illustrated in Fig. S1. Swelling of the PCGO membranes upon exposure to liquid water was monitored by measuring the cross-sectional thickness of the membranes in an optical microscope immediately after and before performing the ion permeation experiments. The increase in thickness after the permeation experiments was found to be <1%, indicating negligible swelling of PCGO membranes. Similarly, the effect of epoxy encapsulation on the $d$ spacing was monitored by measuring the thickness of GO laminates before and after encapsulation. No changes were found. We have also carried out an additional check in which the epoxy encapsulation was removed around one of our GO membranes ($d \approx 7.9$ Å) and X-ray measurements were immediately performed. No change in $d$ (with accuracy of 1-2%) was observed, which confirms the stability of $d$ spacing after the encapsulation procedure.

**Permeation experiments.** All permeation measurements were carried out using the set-up shown in Supplementary Fig. S2, which consists of feed and permeate compartments made from Teflon. PCGO membranes incorporated plastic/metal plates (Supplementary Fig. S1) were clamped between two O-rings and then fixed between the feed and permeate compartments to provide a leak tight environment for the permeation experiments. We filled the compartments with equal volumes (10 mL) of a salt solution (feed) and deionized water (permeate) to avoid any hydrostatic pressure due to different heights of the liquids. Permeation experiments at different temperatures (2–43°C) were performed in a temperature controlled environmental chamber. The measurement-setup, feed and permeate solutions were equilibrated at each temperature before performing the experiment. Magnetic stirring was used in both compartments to avoid concentration polarization effects. Anion and cation concentrations in the permeate compartment caused by diffusion through PCGO membranes were accurately measured using ion chromatography (IC) and inductively coupled plasma atomic emission spectrometry (ICP-AES) techniques[19]. Using the known volume of the permeate compartment, the concentrations allowed us to calculate the amount of ions that diffused into it.

# Supplementary Information

## Tuneable Sieving of Ions Using Graphene Oxide Membranes

1. **Fabrication of physically confined GO (PCGO) membranes**

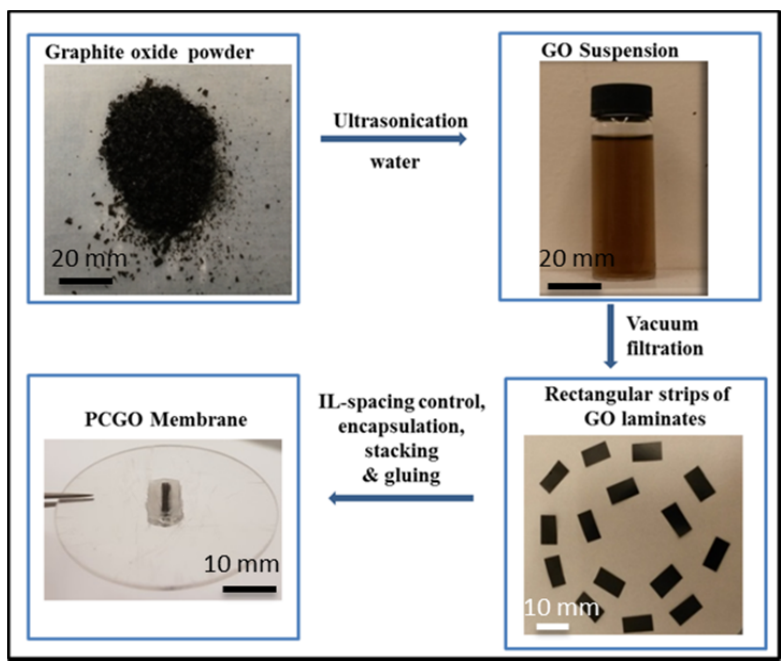

**Supplementary Fig. S1. PCGO membrane fabrication.** Figure illustrating step-by-step procedure in the fabrication of PCGO membrane.

2. **Experimental set-up for permeation experiments**

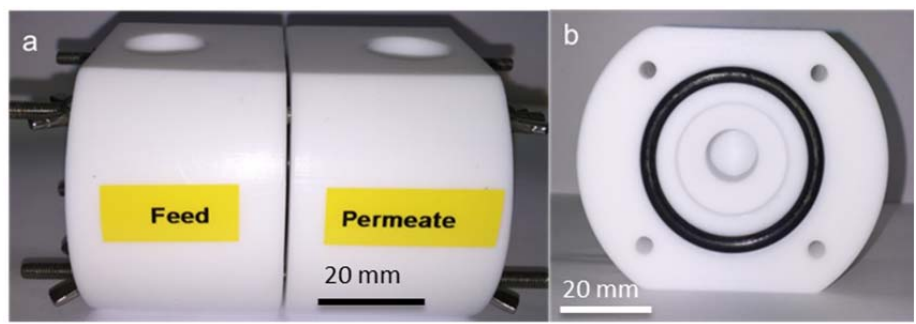

**Supplementary Fig. S2. Permeation set-up.** (a) Experimental set-up showing Teflon made feed and permeate compartments used for the ion permeation experiments. Membranes were clamped between two O-rings and then fixed between feed and permeate compartments to provide a leak tight environment for the permeation experiments. (b) Cross-sectional view of the feed/permeate compartment showing O-ring arrangement for sealing the membranes.

3. **Ion permeation through PCGO membranes**

Ion permeation through PCGO membranes was monitored as a function of concentration gradients and duration of the experiment. As an example, supplementary Fig. S3 shows the results for permeation of $K^+$ and $Cl^-$ ions through PCGO membranes with an interlayer spacing of 9.8 Å. This increases with time in a stoichiometric manner (within our experimental accuracy, as indicated in the figure), to preserve the charge neutrality in both compartments. The slope of such permeation vs time curves gives the permeation rate. As shown in the inset of supplementary Fig. S3, the permeation rate increases linearly with feed concentration, indicating a concentration driven diffusion process[1].

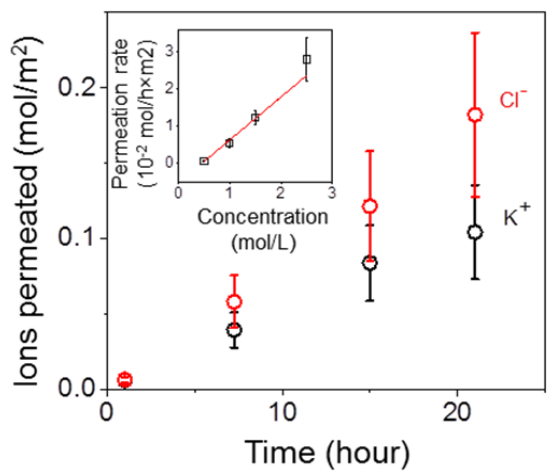

**Supplementary Fig. S3. Ion permeation through PCGO membrane.** Permeation through a PCGO membrane with an interlayer spacing of 9.8 Å from the feed compartment with 1 M aqueous solution of KCl. The error bars indicate our experimental accuracy (~30%) for this particular type of measurements. The inset shows $K^+$ ion permeation rate as a function of concentration of the feed solution. Error bars indicate the standard deviation.

4. **Tested ions and their hydrated diameters**

The hydrated diameters considered for all the ions in Fig. 2 of the main text are obtained from Ref. (2). There are large variations in exact values of hydrated diameters reported in literature[3], due to disparities in the definition and differences in modelling parameters. For example, the reported hydrated diameter of $K^+$ varies from 4 to 6.6 Å and for $Mg^{2+}$ it varies from 6 to 9.4 Å. The chosen values in the main Fig. 2 are 6.6, 7.1, 7.6, 8.2 and 8.5 Å for $K^+$, $Na^+$, $Li^+$, $Ca^{2+}$ and $Mg^{2+}$ respectively. However, irrespective of the chosen hydrated diameter, the absence of a pure size exclusion mechanism in the ion permeation through PCGO membrane is clear. For example, the smallest reported hydrated diameter for $Na^+$ ion is 5.4 Å, so it is not expected to permeate through PCGO membranes with an interlayer spacing smaller than 8.8 Å if the permeation cut-off is dictated by the size exclusion. The observed permeation of $Na^+$ through this membrane confirms that ion permeation through PCGO membranes is not exclusively limited by their hydrated diameter.

5. **Water permeation experiments**

To understand the permeation of water molecules through PCGO membranes we have performed gravimetric measurements[4] and pressure assisted water permeation experiments. Gravimetric measurements were carried out as reported previously[4] inside a glove box environment (< 0.5 ppm of $H_2O$) using a stainless steel container sealed with a PCGO membrane. Air-tight sealing was achieved by fixing the PCGO membrane glued plastic plate to a steel container using two rubber O-rings. In a typical experiment, the weight loss of a water filled container sealed with a PCGO membrane was monitored using computer-controlled balance (Denver Instrument SI-203 with a sensitivity of 1 mg). We have performed the weight loss experiments for the PCGO membranes with interlayer spacing, *d*, of 6.4, 7.4, 7.9, 8.6, 9.0 and 9.8 Å to measure the water permeation rate as a function of interlayer spacing. No noticeable weight loss with an accuracy of 0.2 mg/h×cm$^2$ was observed for the PCGO membranes with 6.4 Å interlayer spacing, indicating that the available free space of ≈ 3 Å is not sufficient for the permeation of water through graphene channels. However, the weight loss rates through PCGO membranes with interlayer spacings of 7.4, 7.9, 8.6, 9.0 and 9.8 Å were measurable and significant: 7.4, 8.8, 10.4, 12.3 and 15.4 mg/h×cm$^2$, giving a water permeance of 3.2, 3.8, 4.5, 5.3 and 6.6 L/h×m$^2$×bar, respectively.

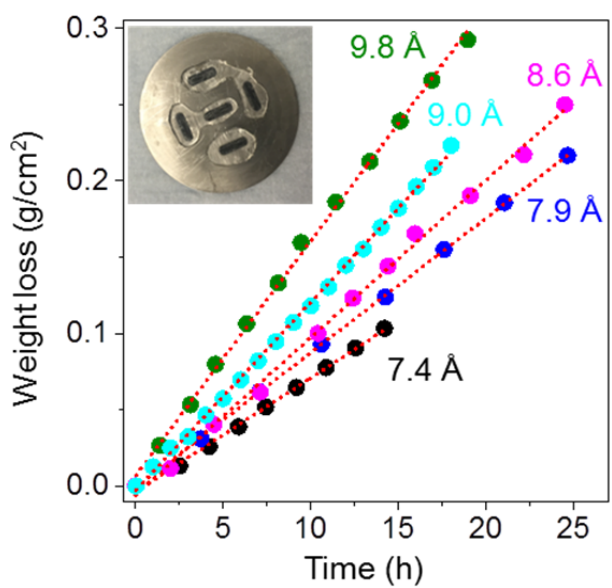

**Supplementary Fig. S4. Water permeation through PCGO membrnaes.** Weight loss for a container sealed with PCGO membrnaes with different interlayer spacing. Inset shows the PCGO membrane sample used for the pressure filtration experiment (diameter of the disc is 51 mm).

In addition to the gravimetric measurements, we have also estimated the rate of liquid water permeation through PCGO membranes with an interlayer spacing of 7.9 Å using a Sterlitech HP4750 stirred cell. As shown in the inset of supplementary Fig. S4, the area of the membrane available for water permeation was increased by gluing multiple stacks of PCGO samples onto a stainless steel plate to collect a measurable amount of permeated water though PCGO membrane. The typical cross-sectional area and permeation length of the PCGO samples in this experiment was 0.3 cm$^2$ and 3 mm, respectively. The PCGO membranes assembly was then fixed inside the stirred cell using a rubber gasket to avoid any possible leakage in the experiment. We have used pure water as a feed solution and collected the water on other side by applying a pressure of 15 bar using a compressed nitrogen gas cylinder. Water permeance was found to be ≈ 0.5-1.0 L/h×m$^2$×bar, which is roughly in agreement with the value obtained from the gravimetric measurements (≈ 4 times smaller). Due to the difficulties of fabricating samples with such large areas for pressure filtration, systematic filtration experiments with salt water were not performed.

Comparison with Hagen-Poiseuille flow equation

Using the standard Hagen-Poiseuille equation with non-slip boundary conditions, we have estimated the water permeation rate through PCGO membranes with different interlayer spacings. Water flow through slit geometry can be described as

$$Q = \frac{1}{12\eta}\frac{\Delta P}{L}\delta^3 W\rho \qquad (S1)$$

where $\eta$ is the viscosity of water (1 mPa.s), $\Delta P$ is driving pressure, $L$ is the permeation length (3 mm), $\delta$ is the effective pore size, $W$ is the lateral width of nanochannels (9 mm) and $\rho$ is the density of water. The water flux through the PCGO membrane can be obtained as $Q \times S$, where $S$ is the area density of nano channels defined as $A/W \times d$, where $A$ is the area and $d$ is the interlayer distance.

For PCGO membranes with an interlayer spacing of 7.4 and 9.8 Å, the estimated water flow rate per cm$^2$ is $\approx$ 2×10$^{-3}$ mg/h and 6×10$^{-3}$ mg/h respectively, which is three orders of magnitude lower than the experimentally observed water flow of 7.4 and 15.4 mg/h respectively. That is, water flow through PCGO membranes with interlayer spacings of 7.4 and 9.8 Å exhibits a flow enhancement, compared to the prediction from the Hagen-Poiseuille equation, by a factor of 4000 and 2000, respectively.

## 6. **Molecular Dynamic Simulations**

Molecular dynamics simulations (MD simulations) were used to calculate the free energy barriers for ions permeating into modelled graphene channels and the diffusion coefficients of the ions inside the channels. All simulations were performed using GROMACS[5], version 5.0.4, in the NVT ensemble at a temperature of 298.15 K, maintained using the Nose-Hoover thermostat[6,7]. The equations of motion were integrated using the leap-frog algorithm[8] with a time-step of 2 fs. The intermolecular potential between particles $i$ and $j$, $V_{ij}$, was evaluated as the sum of a Lennard-Jones 12-6 term and a coulombic term,

$$V_{ij} = 4\varepsilon_{ij}\left[\left(\frac{\sigma_{ij}}{r_{ij}}\right)^{12} - \left(\frac{\sigma_{ij}}{r_{ij}}\right)^{6}\right] + \frac{q_i q_j}{4\pi\varepsilon_0 r_{ij}} \qquad (S2)$$

for which the coulombic term was evaluated using the particle-mesh Ewald[9,10] summation. In Equation S2, $r_{ij}$ is the distance between the two particles with charges $q_i$ and $q_j$ and $\varepsilon_0$ is the vacuum permittivity. In the 12-6 potential, the cross parameters for unlike atoms, $\sigma_{ij}$ and $\varepsilon_{ij}$, were obtained using the Lorentz-Berthelot combining rules,

$$\sigma_{ij} = \frac{(\sigma_i + \sigma_j)}{2} \qquad \text{and} \qquad \varepsilon_{ij} = (\varepsilon_i \varepsilon_j)^{\frac{1}{2}} \qquad (S3)$$

where $\sigma_i$ and $\varepsilon_i$ are the parameters corresponding to an individual atom. Individual carbon atoms in the graphene sheets were modelled as rigid and with zero charge. The parameters for the carbon atoms were obtained from a study in which the water contact angle and adsorption energy were reproduced[11]. The ion parameters were taken from studies in which the hydration free energy and hydrated radius of each ion were calculated and fitted to experimental quantities in bulk solution[12,13]. The original parameterizations of both the carbon and ions were conducted using the SPC/E water model[14] so we have used this model in our simulations. Non-bonded interactions were cutoff for $r_{ij}$ < 1.0 nm. The full set of non-bonded interaction parameters used in the simulations is given in Table S1.

| $i$ | $\sigma_i$ (nm) | $\varepsilon_i$ (kJ mol$^{-1}$) | $q_i$ ($e$) |
|---|---|---|---|
| C | 0.3214 | 0.48990 | 0.000 |
| K$^+$ | 0.4530 | 0.00061 | 1.000 |
| Na$^+$ | 0.3810 | 0.00061 | 1.000 |
| Li$^+$ | 0.2870 | 0.00061 | 1.000 |
| Ca$^{2+}$ | 0.2410 | 0.94000 | 2.000 |
| Mg$^{2+}$ | 0.1630 | 0.59000 | 2.000 |

**Supplementary Table S1.** Non-bonded interaction parameters used in this work.

Free Energy Barriers

The free energy barrier simulations were set up in a similar manner as described in much greater detail in our previous simulations[15]. Briefly, this consists of five layers of graphene sheets, centered in the *x*-direction and stacked parallel in the *z*-direction, with an interlayer spacing of 7, 8, 9, 10 and 11 Å. The interlayer space and adjoining reservoirs were filled with water molecules. A single ion (either Li$^+$, Na$^+$, K$^+$, Mg$^{2+}$ or Ca$^{2+}$) was then swapped for one of the water molecules in the left-hand reservoir to generate the initial configuration (Supplementary Fig. S5).

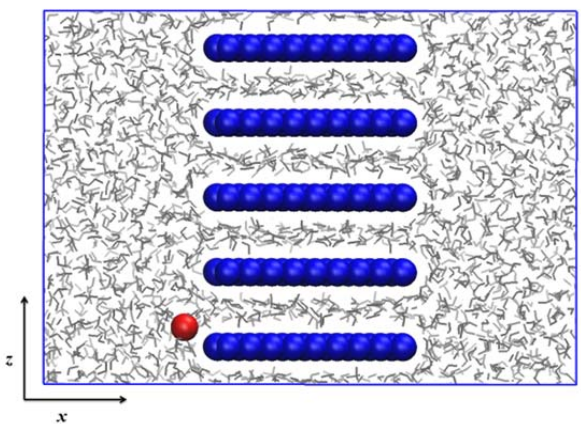

**Supplementary Fig. S5. Free energy barrier simulations.** A snapshot of the simulation cell used in the free energy barrier simulations. The red sphere, blue spheres, and grey lines represent the ion, carbon atoms and water molecules, respectively.

In order to obtain the energy barriers, a potential of mean force (PMF) describing the process of the ion entering the model membrane was generated for every ion and interlayer spacing. This was calculated using an umbrella sampling procedure[16,17] involving 50 separate simulations, spanning the distance from the center of the reservoir ($x$ = 0.1 nm) to the center of the channel ($x$ = 2.5 nm), at 0.05 nm intervals. In each simulation, the position of the ion in the *x* direction was restrained using a harmonic potential with a force constant of 5000 kJ mol$^{-1}$ nm$^{-2}$. After an initial equilibration period of 1 ns, the PMF was generated from the force data obtained in a further 4 ns of simulation time, using the weighted histogram analysis method[18,19]. The maximum energy along the PMF profile is equal to the barrier to permeation. In all cases, the observed barriers are positive, indicating that this process is energetically unfavorable. In general, the barrier height increases as the interlayer spacing decreases and, in the narrowest capillaries, the barriers are considerably larger for divalent ions than monovalent ions. Fig. 3c inset in the main text and supplementary Table S2 show the free energy barriers for every ion obtained for different interlayer spacing.

| Ion | Interlayer Spacing (Å) | | | | |
|---|---|---|---|---|---|
| | 7 | 8 | 9 | 10 | 11 |
| $K^+$ | 27.5(0.6) | 17.4(0.3) | 10.8(0.3) | 5.6(0.2) | 5.6(0.3) |
| $Na^+$ | 22.0(1.1) | 15.9(0.3) | 5.3(0.4) | 5.0(0.3) | 5.3(0.3) |
| $Li^+$ | 24.7(1.3) | 8.5(0.3) | 4.5(0.4) | 3.2(0.3) | 1.8(0.2) |
| $Ca^{2+}$ | 163.5(1.0) | 60.3(0.4) | 3.9(0.3) | 5.5(0.4) | 6.7(0.4) |
| $Mg^{2+}$ | 197.8(2.2) | 44.3(0.5) | 4.6(0.3) | 3.9(0.4) | 5.4(0.4) |

**Supplementary Table S2.** Free energy barriers to ion permeation into graphene capillaries (kJ mol$^{-1}$). The number in brackets is the uncertainty in the size of the barrier.

The observed trends in barrier energy suggest that the size of the barrier is related to the hydration free energy. The higher charge on divalent ions results in stronger electrostatic attraction between the ion and the surrounding water, and the strength of these interactions is reflected in the magnitude of their experimental hydration free energies (see Supplementary Table S3)[20,21]. Hence, ions with the most negative hydration free energies have the largest barriers to permeation, consistent with permeation data obtained experimentally.

| Ion | Hydration free energy (kJ/mol) |
|---|---|
| $K^+$ | -321 |
| $Na^+$ | -405 |
| $Li^+$ | -515 |
| $Ca^{2+}$ | -1592 |
| $Mg^{2+}$ | -1922 |

**Supplementary Table S3.** Experimental hydration free energy of different ions taken from Ref. 2

This ion dehydration effect was further investigated by analyzing the ion hydration numbers in each simulation window along the PMF profile (Supplementary Fig.S6 and S7). The hydration numbers for the first, $n_1$, and second, $n_2$, hydration shells, were calculated by taking the integral at the first and second minima in the ion-water radial distribution function. The Supplementary Fig. S6a. shows that both $n_1$ and $n_2$ decrease as the ions move into a 7 Å channel. Supplementary Fig. S6b. shows that, for $K^+$, $n_1$ decreases to the greatest extent in the narrowest channel. There is a small increase in $n_1$ in the 11 Å channel, relative to bulk solution, and this appears to be because the K-O distance is commensurate with the peaks in the water density profile when $K^+$ is in the center of the channel. We have discussed this observation in our previous work focusing on anion permeation[15]. Typically, $n_1$ and $n_2$ are not integers, because they are averaged over the duration of the simulation and exchange of water molecules between the hydration shells and bulk solution is relatively frequent. However, for the most strongly hydrating ion, $Mg^{2+}$, $n_1$ is always an integer. Supplementary Fig. S7 shows the changes in the first hydration number of $Mg^{2+}$ as the ion enters the channel with interlayer spacing of 7 Å, $n_1 = 6.0$ in bulk solution, $n_1 = 5.0$ at the entrance to the channel, and $n_1 = 4.0$ once in the center of the channel.

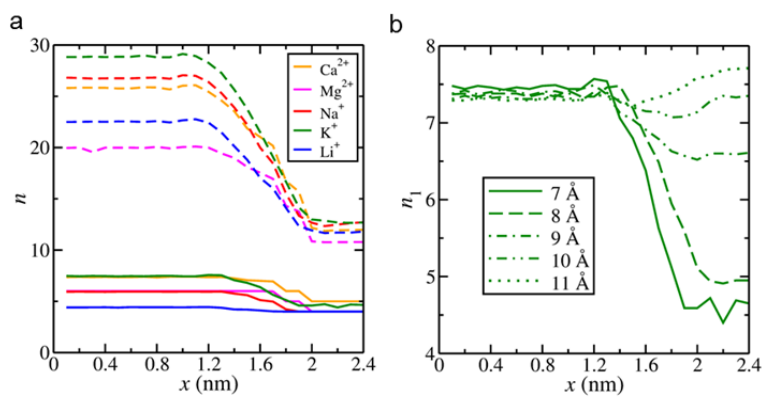

**Supplementary Fig. S6. Ion permeation and ion hydration number** (a) The decrease in $n_1$ (solid line) and $n_2$ (dashed line) as the ions enter a channel with an interlayer spacing of 7 Å. (b) $n_1$ for $K^+$ entering channels with interlayer spacing ranging from 7 to 11 Å.

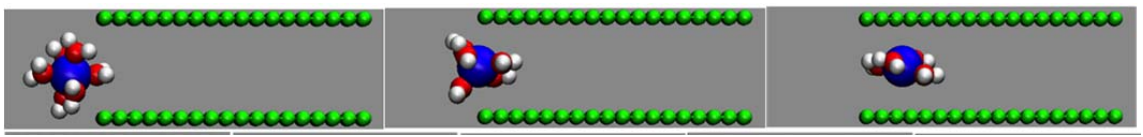

**Supplementary Fig. S7. Dehydration of $Mg^{2+}$.** $Mg^{2+}$ (blue) with the first hydration shell entering the 7 Å graphene channel (green) at $x$ = 1.6, 1.8 and 2.0 nm in the simulation box (left to right).

The primary hydration numbers of ions inside the channel were obtained from the last five simulation windows along the PMF profiles. Supplementary Table S4 shows that $n_1$ decreases with interlayer spacing for all ions. Since the first hydration shell of the $Li^+$ ion is very small, $n_1$ is only reduced slightly from 1.1 nm to 0.7 nm. However, for ions with larger ionic radii the decrease in $n_1$ is more significant. For example, for $K^+$, $n_1$ decreases from 7.7 in a 11 Å channel to 4.7 in a 7 Å channel. Combined with the barriers in Supplementary Table S2, this shows that ions with larger electrostatic interaction with the surrounding water molecules hold more water molecules to the primary hydration shell and shows larger energy barrier for permeation. It is interesting to note that for all of the cations there is a maximum in $n_1$ at some intermediate interlayer spacing. This appears to be the case when the effective interlayer spacing is commensurate with the distance from the ion to the first hydration shell with the ion in the center of the channel. We have also investigated even narrower interlayer spacing (< 0.6 nm) but the channel does not retain any water molecules at this separation so the ions are required to completely dehydrate in order to enter into the membrane in our simulations.

| Ion | Interlayer Spacing (Å) | | | | |
|---|---|---|---|---|---|
| | 7 | 8 | 9 | 10 | 11 |
| $K^+$ | 4.7 | 5.0 | 6.6 | 7.4 | 7.7 |
| $Na^+$ | 4.0 | 4.4 | 5.6 | 5.7 | 5.7 |
| $Li^+$ | 4.0 | 4.0 | 4.4 | 4.2 | 4.2 |
| $Ca^{2+}$ | 5.0 | 7.5 | 7.9 | 7.3 | 7.2 |
| $Mg^{2+}$ | 4.0 | 6.0 | 6.0 | 6.0 | 6.0 |

**Supplementary Table S4.** The number of water molecules in the first hydration shell, $n_1$.

All the above calculations have been performed on pristine graphene capillary. Therefore to clarify the role of oxidized regions on the permeation mechanism we have carried out free energy calculations with a deprotonated OH group (the parameters for the oxygen atom, $q$ = -0.6400 C, $\sigma$ = 0.307 nm, $\varepsilon$ = 0.65 kJ/mol were taken from Ref. 22) attached to the carbon atom at the center of the nanochannel with an interlayer spacing of 8 Å. The resulting free energy barrier for $K^+$ ions turns out to be ~15 kJ/mol similar to that of the pristine channel (17.4 kJ/mol), confirming the dominant importance of the interlayer spacing rather than the chemical functionality for the proposed dehydration mechanism.

Diffusion coefficient of ions inside the sub-nm channels

To calculate the ion's diffusion coefficient, $D$, within the capillary, two graphene sheets with dimensions 6.14 nm x 6.14 nm and interlayer spacing ranging from 7 to 11 Å were used. In this case, unlike the free energy barrier calculations, only one periodic channel was set up, providing an effectively infinitely long 2D capillary for ion diffusion. The density of the water inside the capillary was set up equal to the value obtained from the free energy barrier calculations where the water filled the channel and reached different equilibrium densities as a function of interlayer spacing. After a short equilibration simulation, a single water molecule was exchanged for the ion of interest. Extended simulation runs of 100 ns were used to calculate the mean squared displacement of the ion, and this was used to obtain $D$ the from the Einstein relation

$$\langle |r_i(t_0 + t) - r_i(t_0)|^2 \rangle = 6Dt \tag{S4}$$

where $r_i$ is the position of the particle at time $t_0 + t$ or $t_0$ and the angled brackets denote ensemble averaging. As well as these simulations, we also calculated the diffusion coefficient of $K^+$ in an unconfined box of water molecules (bulk), in order to validate the employed parameters. In this case, the simulation box was cubic, with a side length of 7.5 nm and the simulation was run for 10 ns, using only the final 9 ns in the calculation of $D$. In the unconfined system, we obtained $D = 1.60 \times 10^{-5}$ cm$^2$ s$^{-1}$, which agrees reasonably well with the experimental bulk diffusion coefficient of $1.96 \times 10^{-5}$ cm$^2$ s$^{-1}$ [23]. This shows that our choice of interaction parameters for both the water and $K^+$ ions produce diffusion coefficient in reasonable agreement with experiment, despite dynamic properties not featuring in the original parameterization of the ion – water intermolecular potential.

In the channel, $D$ is reduced relative to the bulk simulation (see supplementary Fig. S8). The difference in diffusion coefficient between bulk and the 8 to 11 Å channel is due to the limited diffusion perpendicular to the graphene sheets. Once the interlayer spacing is reduced below 8 Å, diffusion of $K^+$ is further reduced relative to the bulk; $K^+$ is only able to move within the plane of the single water monolayer at these interlayer spacings. The decrease in $D$ is however modest compared to the decrease in permeation rates observed experimentally. Thus the exponential decrease in the experimental permeation rate with interlayer spacing cannot be explained by the diffusion-limited permeation. This further suggests that the free energy barrier associated with dehydration is the dominant parameter for the ion permeation in our sub-nm capillaries.

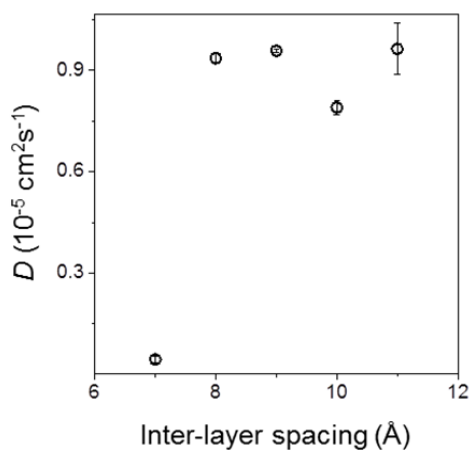

**Supplementary Fig. S8. Ion diffusion through sub-nm channels.** Diffusion coefficient of $K^+$ ion in water for interlayer spacing ranging from 7 Å to 11 Å.

Finally, to completely rule out the diffusion contribution on the experimentally observed ion permeation, we have calculated the capillary diffusion activation energies ($E_a$) of the $K^+$ ion for interlayer distances of 10, 8 and 7 Å by measuring the ion diffusion coefficient at different temperatures ($T$). Supplementary Fig. S9 shows that the diffusion process can be described by an Arrhenius relationship from which we can extract $E_a$. The extracted values of $E_a$ are 15.3 ± 0.4, 12.9 ± 0.3 and 17.7 ± 0.9 kJ mol$^{-1}$ for the interlayer distance 10, 8, and 7 Å, respectively. These values show that $E_a$ is relatively unchanged with interlayer spacing (while the measured barrier increased significantly with decreasing $d$), hence diffusion cannot explain the experimentally observed ion selectivity in sub-nm channels in PCGO membranes.

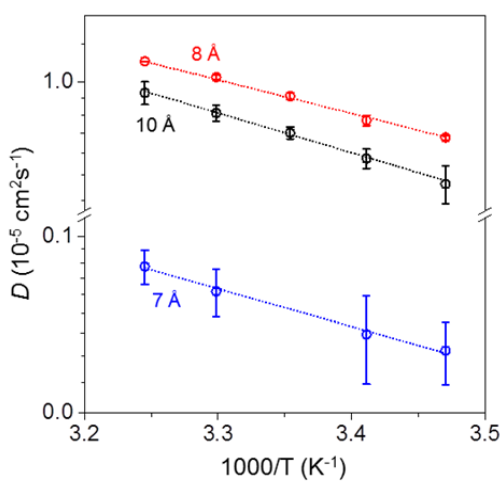

**Supplementary Fig. S9. Diffusion activation energy estimation.** Temperature dependence of $D$ calculated for $K^+$ inside a channel of 10, 8, and 7 Å interlayer spacing (Y-axis - natural log scale). The dashed lines are the best fit to calculate the activation energy.

Permeation rate calculations

To further demonstrate the effect of dehydration on ion permeation rates we have calculated the permeation rate of $K^+$ and $Mg^{2+}$ through a channel with an interlayer spacing of 8 Å. The simulations followed a similar set-up of that used to calculate the energy barrier for ion entry into the channel (Fig S5), except the reservoir of water was larger to allow a concentration of 0.61 mol dm$^{-3}$ of KCl and MgCl$_2$. As at such interlayer spacings there is a large energy barrier associated with the entering of the ions into the channel (Table S2), we do not observe ion permeation over the typical timescale of a simulation. Therefore to calculate the ion permeation rates, a pressure difference of 10 MPa was applied across the simulation cell by adding a constant force on all of the atoms in the simulation box along the direction of the channel, except on those belonging to the graphene sheets[24-27]. During the simulations the temperature was maintained constant at 298.15 K. The interaction parameters for ion, water and graphene atoms were taken the same as the previous simulations. The ion permeation rate was determined by counting the net number of ions that pass from the left to the right reservoir. These simulations were performed for 20 ns, using only the last 15 ns for analysis. The number of $K^+$ and $Mg^{2+}$ ions permeating through the channel has been plotted against time in Fig. S10. This plot clearly shows that the number of $K^+$ ions that permeate through the channel is more than that of $Mg^{2+}$ ions and from this, we have calculated a permeation rate for $K^+$ and $Mg^{2+}$ ions of $1.802(\pm 0.006) \times 10^9$ ions s$^{-1}$ and $0.286(\pm 0.002) \times 10^9$ ions s$^{-1}$.

Not surprisingly, these calculated rates are much higher than the experimental values due to the pressure difference applied. However, they clearly show that permeation into the channel is easier for ions with smaller free energy barrier.

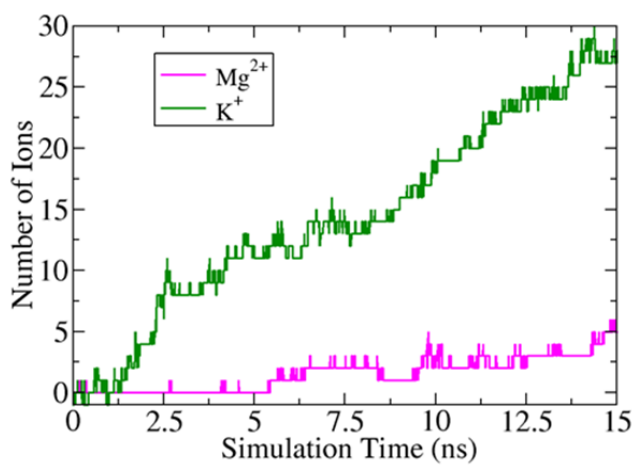

**Supplementary Fig S10. Estimation of Ion permeation rate.** Number of ions permeating through a 8 Å channel during the simulation for $Mg^{2+}$ and $K^+$ with a 10 MPa driving pressure along the channel.

## 7. Swelling-controlled graphene oxide-graphene (GO-Gr) membranes

On its own, water is a poor solvent for the exfoliation of graphite, whereas surfactant-water solutions can exfoliate graphite to produce stable aqueous dispersions of graphene[28]. Graphene oxide (GO) has previously been suggested as a 2D-surfactant to prepare stable dispersions of graphite and carbon nanotubes (CNTs) in water[29,30]. Here, graphene oxide-graphene (GO-Gr) aqueous dispersions were prepared by exfoliating graphite in water using GO as a surfactant. We have prepared four different concentrations of GO-Gr aqueous dispersions by varying the initial weight of bulk graphite with respect to that of graphite oxide. The graphite oxide to graphite weight ratio was maintained as 1:1, 1:2, 1:5 and 1:9, i.e., four different amounts of graphite (0.175 g, 0.35 g, 0.875 g and 1.575 g) were sonicated for 50 hrs in 120 ml of DI water in the presence of 0.175 g graphite oxide. Resulting GO-Gr dispersion was centrifuged twice for 25 mins at 2500 rpm to remove the unexfoliated graphite and unstable aggregates.

Supplementary Fig. S11 shows the optical photograph of GO and GO-Gr aqueous colloidal suspensions of concentration ≈ 0.1 mg/mL, with increasing amounts of exfoliated graphene (from left to right). The pale brown coloured GO suspension gradually turns into black colour as the amount of exfoliated graphene flakes in GO-Gr dispersions increases. AFM images of the GO-Gr dispersion deposited on oxidised silicon wafer show that most of the exfoliated graphene is a few-layers thick (< 5 nm, see supplementary Fig. S11c). GO-Gr membranes were prepared by vacuum filtering each dispersion through an Anodisc alumina membrane filter (25 mm diameter, 0.02 μm pore size) and drying in ambient condition prior to the permeation and X-ray diffraction experiments.

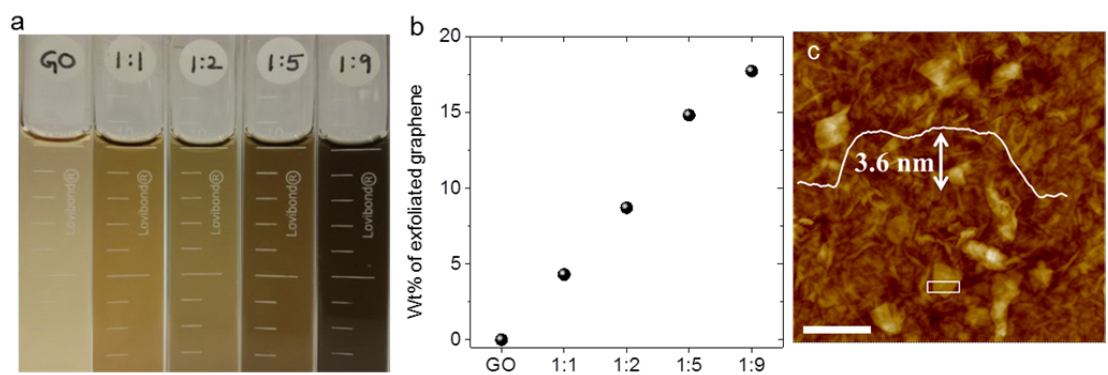

**Supplementary Fig. S11**. **GO-Gr dispersions** (a) Photograph of GO and GO-Gr aqueous colloidal suspensions (concentration ≈ 0.1 mg/mL) with increasing amount of exfoliated graphene (from left to right). (b) Wt% of exfoliated graphene with respect to GO in different GO-Gr membranes. (c) AFM image of GO-Gr thin film deposited on oxidised silicon wafer showing the presence of exfoliated graphene in GO-Gr film. White curve: height profile along the solid rectangle. Scale bar 0.5 μm.

To estimate the concentrations of exfoliated graphene and GO in the GO-Gr dispersions, we measured the weight of the membranes prepared from the known volume of dispersions. Before weighing, the membranes were completely dried in vacuum and the measurements were performed in a glove box to avoid the influence of absorbed water content in the membranes. Supplementary Fig. S11b shows the weight percentage (wt%) of exfoliated graphene flakes calculated from the weighing measurements for different GO-Gr samples. We found that approximately 18 wt%, 15 wt%, 9 wt% and 4.5wt % of exfoliated graphene (with respect to the weight of GO) in the GO-Gr membranes made from the 1:9, 1:5, 1:2 and 1:1 GO-Gr dispersions, respectively. We note that the estimated wt% of exfoliated graphene flakes in GO-Gr membranes represent the lower bound because we assumed that the concentration GO is the same in pristine GO and GO-Gr dispersions. We have also tried to increase the initial GO-graphite ratio above 1:9 but no appreciable change in the concentration of exfoliated graphene was observed in comparison to 1:9 samples.

Characterization of GO-Gr membranes

Supplementary Fig. S12a shows the cross-sectional SEM image of GO-Gr membrane that confirms the laminar structure similar to the pristine GO membranes. In-plane SEM imaging (Supplementary Fig. S12b) suggests a uniform distribution of exfoliated graphene flakes in GO-Gr membrane. Swelling of GO-Gr membranes in liquid water was probed by X-ray diffraction (see main Fig. 3) that revealed significant changes for GO-Gr membranes compared to pristine GO membranes. For example, interlayer spacing of pristine GO, GO-Gr with 4.5, 9, 15 and 18 wt% graphene are 14, 11.9, 11.5, 10.9 and 10.2 Å respectively in liquid water. GO-Gr membranes with 18 wt% graphene exhibited maximum reductions in swelling (≈ 4 Å) and therefore, we have carried out all the ion permeation and forward osmosis experiments with these samples.

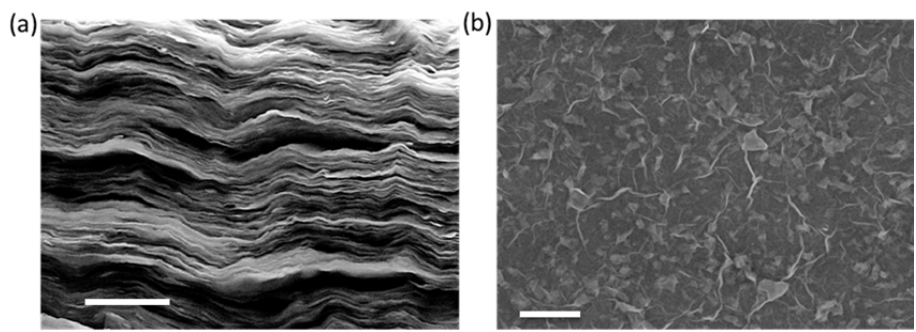

**Supplementary Fig. S12. Electron microscopy on GO-Gr membrnae** (a) Crosssectional and (b) in-plane scanning electron micrograph from the membrane prepared from the 1:9 GO-Gr dispersion. Scale bars are 1 μm.

Pemeation experiments

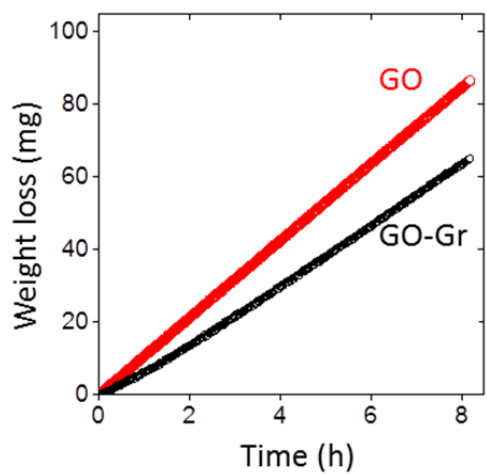

**Supplementary Fig. S13. Water permeation through GO-Gr.** Weight loss for a container filled water sealed with a GO-Gr and a reference GO membrane with a thickness of 5 μm (Area ≈ 0.5 cm$^2$).The weight loss rate for GO and GO-Gr membrane is 10.5 and 8.1 mg/h, respectively.

For all ion permeation experiments we used the same set-up (see Supplementary Fig. S2) as that employed for the PCGO membranes. Ion permeation through GO-Gr membranes was studied by separating the feed and permeate compartment by a 5 μm thick GO-Gr membrane on porous Anodisc alumina support glued onto a plastic disc. The feed and permeate compartments were filled with 1 M aqueous solution of various salts (KCl, NaCl, LiCl and MgCl$_2$) and DI water, respectively. Typically, permeation experiments were carried out for 24 hours and the ion permeation was monitored by ion chromatography (IC) and the inductively coupled plasma optical emission spectrometry (ICP-AES). Similar to the PCGO membranes, ion permeation from feed to permeate compartment through GO-Gr membrane is observed to increase with the duration of experiment and feed concentration. Permeation data for GO-Gr membrane with 18 wt% graphene are shown in Fig. 3 of the main text. Compared to pristine GO membranes, the ion permeation rate for GO-Gr membranes is decreased by two to three orders of magnitude. However, when measured by the gravimetric method, water permeation (see supplementary section 5) only showed an approximately 20% reduction with respect to that of pristine GO (supplementary Fig. S13). The relatively small decrease in water permeation and the large decrease in ion permeation through GO-Gr compared to pristine GO membrane confirm that the permeation mechanism for both PCGO and GO-Gr membranes are similar.

To further understand the liquid water flux and salt rejection properties of GO-Gr membranes, we have performed forward osmosis (FO)[31,32] experiments. FO is relatively a new alternative technology to the conventional pressure-driven reverse osmosis (RO) membrane process, where hydraulic pressure is not required for its operation[31,32]. In FO, a concentrated solution of a salt or other molecules (draw solution) is used to generate high osmotic pressure, which pulls the water molecules across a semi-permeable membrane from the low-concentration salt solution (feed solution), effectively filtering the feed water. The draw solute can then be separated from the diluted draw solution to produce clean water. FO has many advantages over conventional RO such as high energy efficiency and low fouling and is considered to be an attractive emerging technology for desalination. The absence of hydraulic pressure in FO makes it highly suitable to evaluate GO-Gr membranes as they have relatively weak mechanical strength. Here, we have performed FO by filling equal volumes (25 mL) of 0.1 M NaCl feed solution and 3 M sucrose draw solution in the feed and permeate compartments, respectively, separated by a GO-Gr membrane (5 μm thick and 0.5 cm$^2$ area). Nearly 3 M differential concentration leads to a ≈ 75 bar osmotic pressure gradient, which draws water molecules from the NaCl compartment to the sucrose compartment. The amount of water permeation was reflected in the height of sucrose column in the permeate compartment. The observed 0.8 mL increase in the column height over 30 hours corresponds to a water flux of around 0.5 L/m$^2$×h. Salt rejection for GO-Gr membrane was estimated by measuring the amount of NaCl in the draw solution. The salt rejection rate was estimated as 1-$C_d$/$C_f$, where $C_d$ and $C_f$ are the concentrations of NaCl in the draw solution and the feed side, respectively. This yielded a rejection rate of ≈ 97%. For comparison, we have also performed similar FO experiments with pristine GO membranes and the obtained water flux and salt rejection are found to be 0.6 L/m$^2$×h and 60%, respectively. We note that the water flux through GO-Gr membranes is lower than typical FO membranes however it can be improved effectively by decreasing the thickness of GO-Gr membranes. For example, decreasing the GO-Gr membrane thickness from 5 μm to 1 μm yielded the water flux of 2.5 L/m$^2$×h with 94% salt rejection.

Supplementary References